\documentclass[twocolumn,aps,amsmath,amssymb,floatfix,superscriptaddress,prb,longbibliography]{revtex4-1}
\usepackage{graphicx}
\usepackage{dcolumn}
\usepackage{bm}
\usepackage[normalem]{ulem}
\usepackage{xspace}
\usepackage{hyperref}
\usepackage{soul}
\usepackage{color}
\usepackage[flushleft]{threeparttable}
\usepackage{mhchem}
\def\new{\color{black}}

\def\NNBO{Na$_3$Ni$_2$BiO$_6$\xspace}

\begin{document}



\title{Observation of a 1/3 Magnetisation Plateau Phase {\new as Evidence for }the Kitaev Interaction in a Honeycomb-Lattice Antiferromagnet}

\author{Yanyan~Shangguan}
\altaffiliation{These authors contributed equally to the work.}
\author{Song~Bao}
\altaffiliation{These authors contributed equally to the work.}
\affiliation{National Laboratory of Solid State Microstructures and Department of Physics, Nanjing University, Nanjing 210093, China}
\author{Zhao-Yang~Dong}
\altaffiliation{These authors contributed equally to the work.}
\affiliation{Department of Applied Physics, Nanjing University of Science and Technology, Nanjing 210094, China}
\author{Ning~Xi}
\altaffiliation{These authors contributed equally to the work.}
\affiliation{CAS Key Laboratory of Theoretical Physics, Institute of Theoretical Physics, Chinese Academy of Sciences, Beijing 100190, China}
\altaffiliation{These authors contributed equally to the work.}
\author{Yi-Peng~Gao}
\affiliation{National Laboratory of Solid State Microstructures and Department of Physics, Nanjing University, Nanjing 210093, China}
\author{Zhen~Ma}
\affiliation{Institute of Applied Physics, Hubei Normal University, Huangshi 435002, China}
\author{Wei~Wang}
\affiliation{School of Science, Nanjing University of Posts and Telecommunications, Nanjing 210023, China}
\author{Zhongyuan~Qi}
\affiliation{Center of Materials Science and Optoelectronics Engineering, College of Materials Science and Optoelectronic Technology, University of Chinese Academy of Sciences, Beijing 100049, China}
\author{Shuai~Zhang}
\author{Zhentao~Huang}
\author{Junbo~Liao}
\author{Xiaoxue~Zhao}
\author{Bo~Zhang}
\author{Shufan~Cheng}
\author{Hao~Xu}
\affiliation{National Laboratory of Solid State Microstructures and Department of Physics, Nanjing University, Nanjing 210093, China}
\author{Dehong~Yu}
\author{Richard~A.~Mole}
\affiliation{Australian Nuclear Science and Technology Organisation, Lucas Heights, New South Wales 2234, Australia}
\author{Naoki~Murai}
\author{Seiko~Ohira-Kawamura}
\affiliation{J-PARC Center, Japan Atomic Energy Agency (JAEA), Tokai, Ibaraki 319-1195, Japan}
\author{Lunhua~He}
\affiliation{Beijing National Laboratory for Condensed Matter Physics, Institute of Physics, Chinese Academy of Sciences, Beijing 100190, China}
\affiliation{Spallation Neutron Source Science Center, Dongguan, 523803, China}
\affiliation{Institute of High Energy Physics, Chinese Academy of Sciences, Beijing, 100049, China}
\author{Jiazheng~Hao}
\affiliation{Spallation Neutron Source Science Center, Dongguan, 523803, China}
\affiliation{Institute of High Energy Physics, Chinese Academy of Sciences, Beijing, 100049, China}
\author{Qing-Bo~Yan}
\affiliation{Center of Materials Science and Optoelectronics Engineering, College of Materials Science and Optoelectronic Technology, University of Chinese Academy of Sciences, Beijing 100049, China}
\author{Fengqi~Song}
\affiliation{National Laboratory of Solid State Microstructures and Department of Physics, Nanjing University, Nanjing 210093, China}
\affiliation{Collaborative Innovation center of Advanced Microstructures, Nanjing University, Nanjing 210093, China}
\author{Wei~Li}
\email{w.li@itp.ac.cn}
\affiliation{CAS Key Laboratory of Theoretical Physics, Institute of Theoretical Physics, Chinese Academy of Sciences, Beijing 100190, China}
\author{Shun-Li~Yu}
\email{slyu@nju.edu.cn}
\author{Jian-Xin Li}
\email{jxli@nju.edu.cn}
\author{Jinsheng Wen}
\altaffiliation{jwen@nju.edu.cn}
\affiliation{National Laboratory of Solid State Microstructures and Department of Physics, Nanjing University, Nanjing 210093, China}
\affiliation{Collaborative Innovation center of Advanced Microstructures, Nanjing University, Nanjing 210093, China}

\begin{abstract}
\noindent{\bf Fractional magnetisation plateaus, in which the magnetisation is pinned at a fraction of its saturated value within a range of external magnetic field, are spectacular macroscopic manifestations of the collective quantum behaviours. One prominent example of the plateau phase is found in spin-1/2 triangular-lattice antiferromagnets featuring strong geometrical frustration, and is often interpreted as quantum-fluctuation-stabilised state in magnetic field via the ``order-by-disorder" mechanism. Here, we observe an unprecedented 1/3 magnetisation plateau between 5.2 and 7.4~T at 2~K in a spin-1 antiferromagnet \NNBO with a honeycomb lattice, where conventionally no geometrical frustration is anticipated. By carrying out elastic neutron scattering measurements, we {\new propose} the spin structure of the plateau phase to be an unusual partial spin-flop ferrimagnetic order, transitioning from the zigzag antiferromagnetic order in zero field. Our theoretical calculations show that the plateau phase is stabilised by the bond-anisotropic Kitaev interaction. These results provide a new paradigm for the exploration of rich quantum phases in frustrated magnets and exotic Kitaev physics in high-spin systems.}
\end{abstract}
\maketitle

\noindent{\bf Main}\\
Frustration, which describes the situation that competing magnetic exchange interactions cannot be satisfied simultaneously, plays an essential role in quantum magnets\cite{nature464_199}. The frustration-induced quantum fluctuation could avoid the formation of ordered magnetic ground states, and lead to magnetically disordered phases such as the quantum spin liquids\cite{npjqm4_12,Broholmeaay0668}. On the other hand, the quantum fluctuation, which is represented by the zero-point oscillation energy in the spin-wave theory, can lift the degeneracy of the ground state and select a specific spin state within a finite range of external magnetic field; this gives rise to an exotic magnetisation plateau phase in which the magnetisation is a fraction of its saturation value, understood as the ``order-by-disorder" mechanism\cite{Chubukov_1991,Honecker_1999,0034-4885-78-5-052502,PhysRevLett.85.3269,kawamura1985phase,PhysRevLett.62.2056,PhysRevLett.102.137201,PhysRevB.87.060407,PhysRevLett.112.127203}. Such a quantization of a macroscopic physical quantity in a range of magnetic field is a spectacular demonstration of the macroscopic quantum phenomena.

\begin{figure*}[htb]
\centering
\includegraphics[width=6.8in]{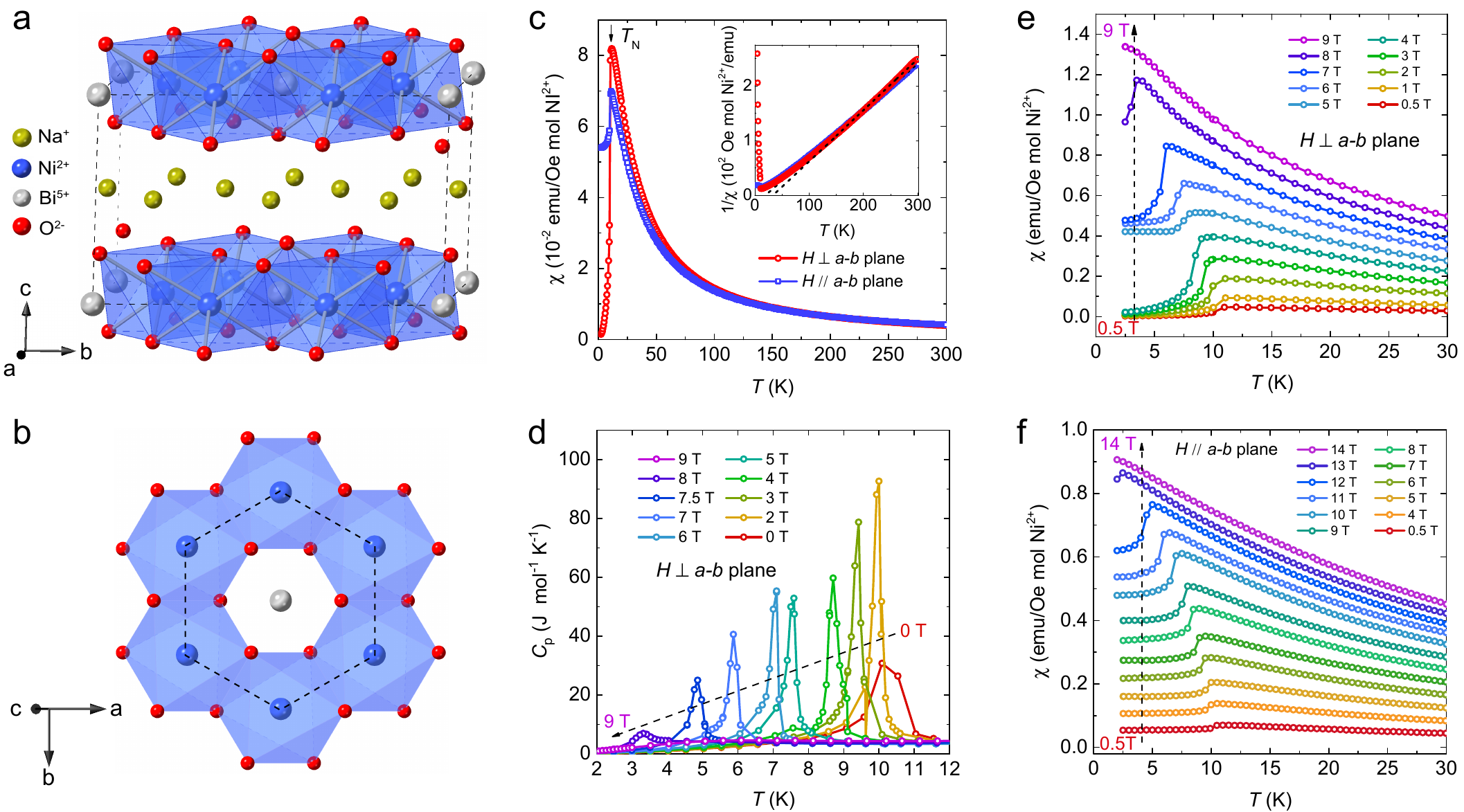}
\caption{\label{fig1}{{\bf Structure and anisotropic antiferromagnetic order of \NNBO single crystals.} {\bf a}, Schematic crystal structure of \NNBO. {\bf b}, Top view of the honeycomb layer of NiO$_6$ octahedra. Dashed lines in {\bf a} represent the crystalline unit cell and those in {\bf b} represent the honeycomb unit formed by the Ni$^{2+}$ ions. {\bf c}, Magnetic susceptibility for field $\mu_{0}H=1$~T applied parallel and perpendicular to the $a$-$b$ plane for a \NNBO single crystal. The inset shows the inverse susceptibilities and accompanying Curie-Weiss fits. {\bf d}, Specific heat in various magnetic fields up to 9~T applied perpendicular to the $a$-$b$ plane, showing that the transition temperature is gradually suppressed with increasing field. {\bf e,f}, Magnetic susceptibility in a series of magnetic fields up to 14~T applied perpendicular and parallel to the $a$-$b$ plane, respectively. The results show that the antiferromagnetic order is highly anisotropic in and out of the $a$-$b$ plane. Data are offset vertically for clarity. Errors represent one standard deviation throughout the work. In Figs.~\ref{fig1}, \ref{fig2}, and \ref{fig4}, they are smaller than the symbols.}}
\end{figure*}

One prominent example is the 1/3 magnetisation plateau, which has been theoretically predicted and experimentally observed in spin-1/2 equilateral-triangular-lattice antiferromagnets, where the spin is small and the frustration
due to the geometrical configuration is strong\cite{Schotte_1994,PhysRevB.67.104431,PhysRevB.76.060406,PhysRevLett.102.257201,prl108_057205,prl109_267206,PhysRevLett.110.267201,kamiya2018nature}. The discussions have been extended to other triangular-lattice systems with higher spins\cite{ToshiyaInami1996,doi:10.1143/JPSJ.80.093702,PhysRevLett.109.257205}, and other two-dimensional systems such as kagome\cite{PhysRevLett.88.057204,PhysRevLett.97.067202,nc4_2287}, and square-lattice antiferromagnets\cite{LOZOVIK1993873,PhysRevLett.82.3168,doi:10.1126/science.1075045} with geometrical and exchange frustrations, respectively. There are also some attempts in honeycomb-like antiferromagnets\cite{PhysRevB.93.094420}, but only some traces have been observed in polycrystalline samples under extremely high fields\cite{doi:10.7566/JPSJ.88.013703}. Since frustration is a prerequisite for the fractional magnetisation plateau phase, whether it will occur in a genuine honeycomb lattice where geometrical frustration as those in triangular and kagome lattices is absent, and how to understand it if it does occur, remain outstanding questions.

Here, we report comprehensive thermodynamic and neutron scattering measurements on high-quality single crystals of the spin-1 honeycomb-lattice antiferromagnet \NNBO~(Ref.~\cite{doi:10.1021/ic402131e}).
We show that the magnetisation curve has a definite plateau at 1/3 of the saturation magnetisation between 5.2 and 7.4~T at 2~K. From our neutron scattering measurements, we obtain complete contour maps for the magnetic Bragg peaks in the ($H,\,K,\,0$) plane in zero and 6.6-T field, the latter of which keeps the system in the plateau phase. {By comparing experimental results with calculated magnetic structure factors for all the possible spin states within reasonably large 24 lattice sites}, we {\new propose the microscopic magnetic configuration of the 1/3 magnetisation plateau phase to be a zero-up-zero-down-up-up~($\circ$$\uparrow
$$\circ$$\downarrow\uparrow\uparrow$) ferrimagnetic state.} Such a state is a transition from the zigzag order ground state induced by the partial spin-flop process, where two of the six spins in an enlarged magnetic unit flop onto the honeycomb plane and exhibit zero magnetic moment along the out-of-plane direction.  Based on these results, we establish a magnetic phase diagram including a salient 1/3 plateau phase for \NNBO. Our density-functional-theory (DFT) and tensor-network calculations show that a minimal model with Heisenberg exchange couplings $J$, a bond-dependent anisotropic Kitaev interaction $K$ and  a single-ion anisotropy term $D$ can well explain the experimental observations. In particular, the Kitaev interaction which was proposed earlier for this and other materials such as $\alpha$-RuCl$_3$ and iridates\cite{PhysRevLett.123.037203,0953-8984-29-49-493002,nrp1_264,PhysRevB.94.064435}, leads to the exchange frustration and stabilises the 1/3 plateau phase. These results suggest \NNBO to be a fertile ground to investigate the quantum physics in frustrated magnets on a honeycomb lattice.

\begin{figure}[htb]
\centering
\includegraphics[width=3.5in]{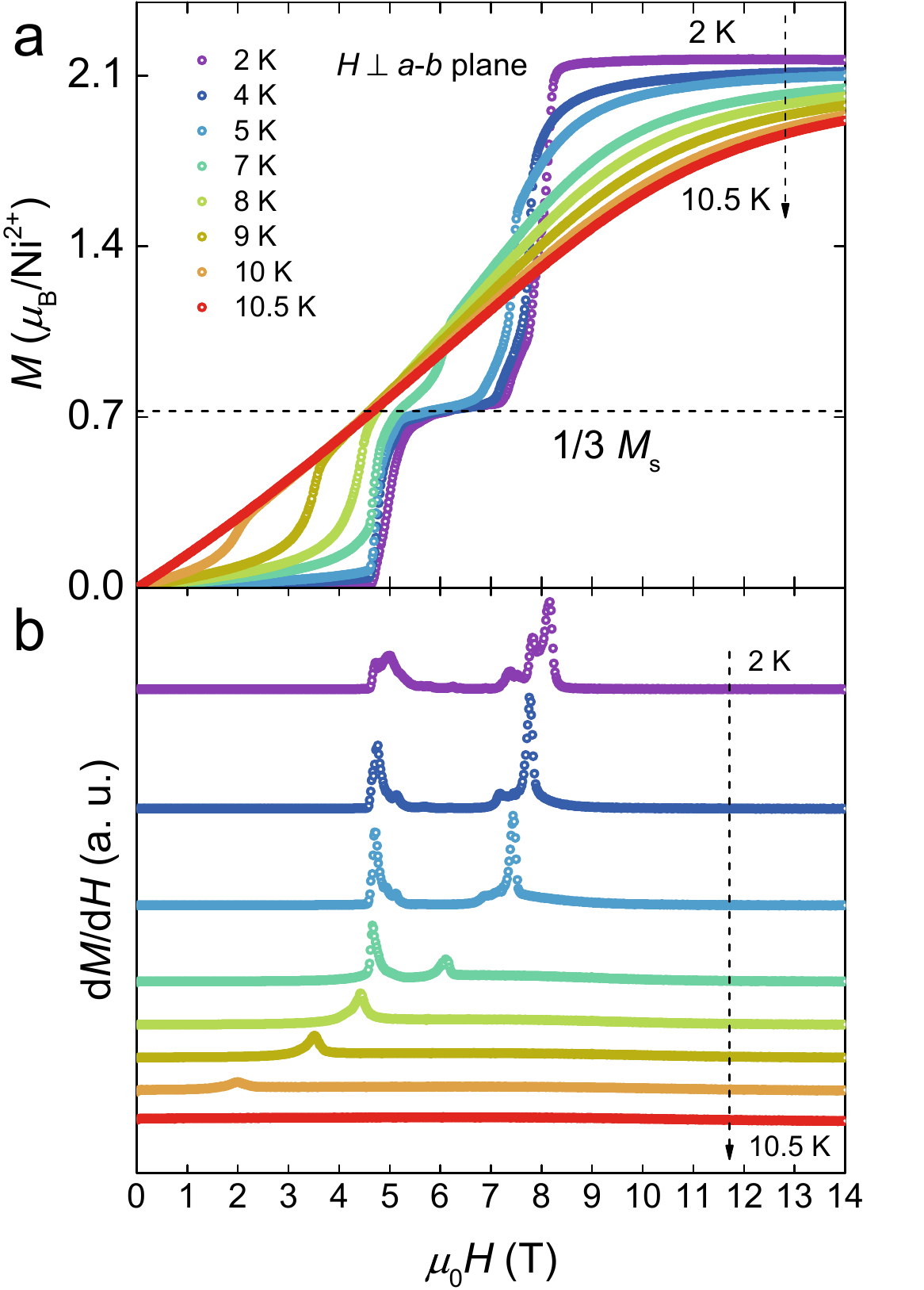}
\caption{\label{fig2}{{\bf 1/3 magnetisation plateau of \NNBO.} {\bf a}, magnetisation as a function of magnetic field applied perpendicular to the $a$-$b$ plane for a \NNBO single crystal. The dotted line indicates 1/3 of the saturated magnetic moment, signifying a fractional magnetisation plateau. {\bf b}, Derivative of the magnetisation in {\bf a} versus magnetic field. Data are offset vertically for clarity.}}
\end{figure}

\medskip
\noindent{\bf Anisotropic antiferromagnetic order}\\
As shown in Fig.~\ref{fig1}a,b, \NNBO has a quasi-two-dimensional structure where the spin-1 Ni$^{2+}$ ions form the honeycomb lattice on the $a$-$b$ plane\cite{doi:10.1021/ic402131e}. The long-range antiferromagnetic order has an onset at the N\'eel temperature $T_{\rm N} \sim$10~K~(Fig.~\ref{fig1}c).
Upon cooling below $T_{\rm N}$, we find that the susceptibility drops rapidly to zero for $H \perp a$-$b$ plane. On the other hand, the susceptibility for $H \parallel a$-$b$ plane drops only slightly. These results indicate that \NNBO is a collinear antiferromagnet with strong anisotropy. In the inset of Fig.~\ref{fig1}c, we show the inverse susceptibilities and Curie-Weiss fits, from which we obtain the effective magnetic moments, $\mu_{\rm{eff}}=2.92(3)$ and 3.05(6)~$\mu_{\rm{B}}$/Ni$^{2+}$ for field applied perpendicular and parallel to $a$-$b$ plane, respectively. These values are close to the spin-only effective moment $\mu_{\rm{eff}}=2.828$~$\mu_{\rm{B}}$/Ni$^{2+}$ with $S = 1$. The Curie-Weiss temperature $\Theta_{\rm CW}$ are 18.87(7) and 33.55(6)~K for fields applied in and out of plane, respectively.


In Fig.~\ref{fig1}e,f, we plot the temperature dependence of the magnetic susceptibility under various magnetic fields applied perpendicular and parallel to the $a$-$b$ plane, respectively. In both cases, the $T_{\rm N}$ is reduced with increasing field, but the trend is more moderate for $H\parallel$ $a$-$b$ plane. For $H \perp a$-$b$ plane, the transition disappears at $\mu_{0}H=9$~T, while for parallel field of 9~T, the system has a $T_{\rm N}$ of 8~K and the transition disappears up to 14~T, indicating strong anisotropy along the out-of-plane direction. Notably, in Fig.~\ref{fig1}e, for $\mu_0H_{\perp}\leqslant4$~T, the susceptibilities finally drop to zero as the temperature decreases, but increase by two successive large steps at 4-5~T and 7-8~T at low temperatures. This can be attributed to the presence of a 1/3 magnetisation plateau phase under perpendicular magnetic fields, as discussed in details below. The susceptibility increases progressively with field for $H\parallel$ $a$-$b$ plane below 13~T, unlike the case for $H\perp$ $a$-$b$ plane. In addition, Fig.~\ref{fig1}d shows the temperature dependence of the specific heat, measured with $H \perp a$-$b$ plane. In zero field, we obtain a $T_{\rm N}$ of $\sim$10.1~K, similar to that in Ref.~\cite{doi:10.1021/ic402131e}. The $T_{\rm N}$ progressively decreases with increasing magnetic field and eventually vanishes at 9~T, consistent with the magnetic susceptibility data.

\begin{figure*}[htb]
\centering
\includegraphics[width=6.0in]{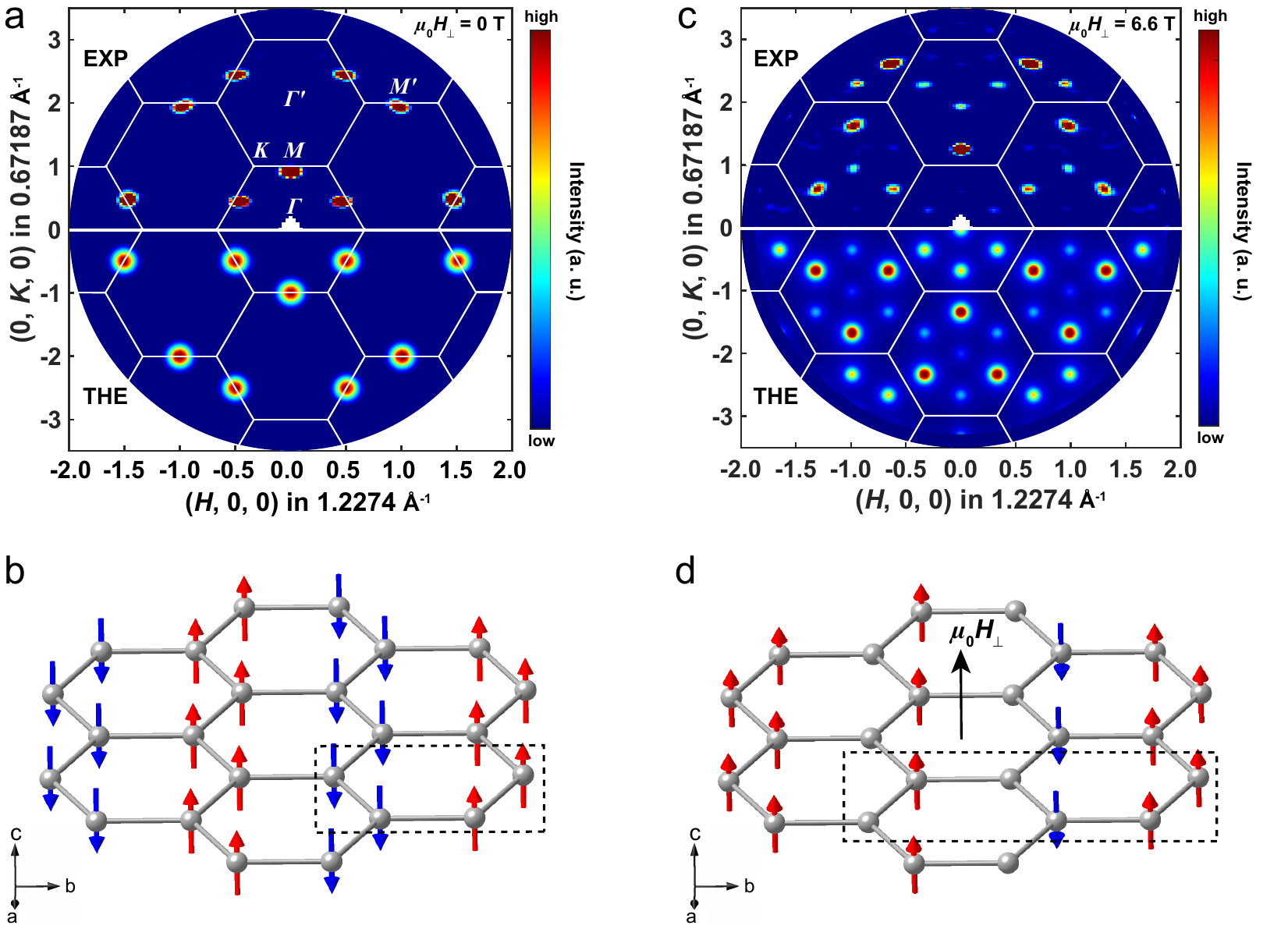}
\caption{\label{fig3}{{\bf Zigzag magnetic structure in zero field, and partial spin-flop structure of the 1/3 magnetisation plateau phase under field.} Top panels of {\bf a,c} show elastic neutron scattering results in the ($H$,\,$K$,\,0) plane measured with $E_{\rm i}= 3.70$~meV on PELICAN under magnetic field of $\mu_0H_{\perp}=0$ and 6.6~T, respectively. The results are obtained by subtracting the 80-K data from 1.5-K data, and then are corrected by the magnetic form factors of Ni$^{2+}$.  Solid lines mark the Brillouin zones. {\bf b,d}, Spin configurations of the zero-field ground state and the 1/3 magnetisation plateau phase, respectively. The former is the zigzag antiferromagnetic order with magnetic moments aligned along the $c$ axis. Spins align ferromagnetically within a zigzag chain, and antiferromagnetically between chains, in an up-up-down-down manner~($\uparrow\uparrow\downarrow\downarrow$). {\new The 1/3 magnetisation plateau phase has an exotic ferrimagnetic order with spins arranged in a zero-up-zero-down-up-up fashion~($\circ$$\uparrow$$\circ$$\downarrow$$\uparrow\uparrow$)}, induced by a partial spin-flop transition, in which two of the six spins in an enlarged magnetic unit flop onto the honeycomb plane and exhibit zero magnetic moment along the out-of-plane direction. Dotted boxes represent the magnetic unit with four spins in {\bf b} and six spins in {\bf d}, for the $\downarrow\downarrow\uparrow\uparrow$ and {\new $\circ$$\uparrow$$\circ$$\downarrow$$\uparrow\uparrow$} structures, respectively. In {\bf d}, the up and down spins are aligned along $c^*$ parallel to the magnetic field. Calculated magnetic structure factors corresponding to the structures in {\bf b,d} are plotted in the bottom panels of {\bf a,c}, respectively. In the calculations, the intensity is symmetrised with respect to the three different magnetic domains. ``EXP" and ``THE" stand for ``Experiment" and ``Theory", respectively.}}
\end{figure*}

\medskip
\noindent{\bf 1/3 magnetisation plateau}\\
Figure~\ref{fig2}a shows magnetisation results as a function of field at different temperatures for $H\perp$ $a$-$b$ plane. There are three prominent plateaus of the magnetisation curves measured at low temperatures. Taking 2~K as an example, first, a zero-magnetisation plateau persists up to $\sim$4.6~T, which corresponds to a spin gap and is usually expected in systems with an Ising-like anisotropy\cite{Chubukov_1991,Honecker_1999,0034-4885-78-5-052502,PhysRevLett.85.3269}. Second, the magnetisation is pinned at $\sim$0.72(5)~$\mu_{\rm{B}}$/Ni$^{2+}$ with fields ranging from 5.2 to 7.4~T, giving rise to a quantized magnetisation plateau. Finally, the magnetisation saturated completely at 8.3~T with the saturation ferromagnetic moment of $\sim$2.16(5)~$\mu_{\rm B}$/Ni$^{2+}$, which is within the reasonable range for the estimated saturated moment of $M_{s}=g\mu_{\rm{B}}S=2$~$\mu_{\rm B}$/Ni$^{2+}$. Evidently, the observed fractional magnetisation plateau is close to 1/3 of the saturated ferromagnetic moment. Upon warming, the fractional plateau gradually diminishes, indicating that it is the quantum fluctuation instead of the thermal fluctuation that stabilises the plateau\cite{Chubukov_1991,Honecker_1999,0034-4885-78-5-052502,PhysRevLett.85.3269,kawamura1985phase,PhysRevLett.62.2056,PhysRevLett.102.137201,PhysRevB.87.060407,PhysRevLett.112.127203}. With increasing temperature, the saturation field increases, while the saturation moment decreases and the saturation plateau shrinks.

To explicitly elucidate the magnetisation process for $H \perp a$-$b$ plane, we plot the differential of magnetisation ($dM/dH$) versus magnetic field at different temperatures  in Fig.~\ref{fig2}b. At 2~K, we find the curve displays two sharp peaks at 4.7 and 8.2~T, which represent the lower- and upper-bound fields of the 1/3 magnetisation plateau state. As the temperature increases, while the lower-bound field remains almost unchanged, the upper-bound field is reduced quickly, so that the plateau shrinks. The robustness of the lower-bound field is a probable nontrivial signature of the quantum nature of the transition wherein. At 8~K and above, the two peaks merge into one. The single peak corresponds to the onset field of the spin-flop transition, which can also be visualized from the magnetisation curves shown in Fig.~\ref{fig2}a. Eventually, all peaks disappear at 10.5~K. Such a transition temperature agrees well with the $T_{\rm N}$ obtained from the specific heat measurements.

Combining the data in Fig.~\ref{fig2}a,b, we estimate a prominent 1/3 magnetisation plateau located between 5.2 and 7.4~T at 2~K, which shrinks with increasing temperature and survives up to $\sim$8~K. Note that this is narrower than the range determined by the upper- and lower-bound fields considering the finite transition width. In contrast, as illustrated in Extended Data Fig.~1a,b, there is no similar 1/3 magnetisation plateau observed for field applied within the $a$-$b$ plane up to 14~T due to the easy-axis anisotropy. However, there is a step-like transition in high fields and low temperatures, which corresponds to the spin-flop transition induced by the transverse magnetic field\cite{Chubukov_1991,PhysRevB.67.104431}.



\medskip
\noindent{\bf Partial spin-flop structure of the 1/3 plateau phase}\\
To shed light on the microscopic mechanism responsible for the above 1/3 magnetisation plateau phase, we next explore the magnetic structures of the ground state in zero field, and especially the 1/3 magnetisation plateau phase under intermediate fields, using elastic neutron scattering on a spectrometer PELICAN on high-quality single-crystalline samples. Top panels of Fig.~\ref{fig3}a,c display contour maps of elastic scattering in the ($H,\,K,\,0$) plane measured with $\mu_{0}H=0$ and 7~T, respectively, with field applied along the $c$ axis. In both panels, the scattering data are plotted by subtracting the 80-K data (well above the $T_{\rm N}$) from the 1.5-K data (below $T_{\rm N}$) so that we can obtain pure magnetic signals by eliminating the lattice contributions (See Extended Data Fig.~2 for the raw data at 1.5 and 80~K). In zero field, we find various magnetic Bragg peaks at ($\pm0.5$,\,$\pm0.5$,\,0), ($\pm1.5$,\,$\pm0.5$,\,0), (0,\,$\pm1$,\,0), ($\pm1$,\,$\pm2$,\,0), and ($\pm0.5$,\,$\pm2.5$,\,0). From Fig.~\ref{fig3}a and Extended Data Fig.~2c, we do not  observe any extra magnetic or structural Bragg peaks due to the inter-occupancy of the Ni and Bi atoms\cite{doi:10.1021/ic402131e}, so we believe that the honeycomb layer formed by the Ni atoms remains intact. The appearance of magnetic Bragg peaks at the Brillouin zone boundary $M$ points is consistent with the zigzag antiferromagnetic order suggested by a preliminary neutron powder diffraction study\cite{doi:10.1021/ic402131e}. In fact, our first-principles calculations also show that the zigzag order state has the lowest energy~(Extended Data Table~2). To gain further insights into the magnetic structure, we have carried out new neutron diffraction measurements on the single crystals on another spectrometer AMATERAS, which has a much larger ${\bf Q}$ coverage due to the multiple-incident-energy option and finer resolutions. The results allow us to perform refinements from which we obtain the moment direction to be $19.7\pm2.6^\circ$ off the $c^*$, close to the \textit{c} axis (see Methods and Extended Data Fig.~3a). This angle is further confirmed by our DFT calculations shown in Extended Data Fig.~3b. The zigzag magnetic order is depicted in Fig.~\ref{fig3}b, where the spins align ferromagnetically along the \textit{c} axis within a zigzag chain, and antiferromagnetically between the chains, exhibiting an up-up-down-down order~($\uparrow\uparrow\downarrow\downarrow$). These results are consistent with those in Ref.~\cite{doi:10.1021/ic402131e}. We calculate the magnetic structure factors for this structure and the results are shown in the bottom panel of Fig.~\ref{fig3}a. It is clear that the calculations agree with the experimental data well. From these results, we are able to pin down the magnetic structure to be the zigzag antiferromagnetic order as shown in Fig,~\ref{fig3}b.

At $\mu_{0}H=7$~T, corresponding to $\mu_{\perp}H = 6.6$~T along $c^*$, which keeps the system in the 1/3 magnetisation plateau phase, the scattering patterns become more complicated. In the first Brillouin zone, there are actually seven magnetic Bragg peaks, with one in the zone centre (and thus cannot be observed experimentally due to the angle restriction), and another six around the centres of the triangles formed by the $\mit\Gamma$ and $K$ points with equivalent intensity. In the second Brillouin zone, we can observe seven peaks at positions identical to those in the first Brillouin zone. But intriguingly, the six peaks around the triangle centres can be divided into two groups rotated by 180$^\circ$ with dramatically different intensities. Furthermore, the stronger peaks in the second Brillouin zone appear to be more intense than those in the first Brillouin zone. Based on the zigzag ground state in zero field, we have exhausted all the possible magnetic structures within reasonably large lattice sites (up to 24) for the 1/3 plateau phase in Extended Data Fig.~4, and calculated the corresponding magnetic structure factors in Extended Data Fig.~5. In the calculations, we have tuned the angle for the out-of-plane moments and found the results are hardly affected. Therefore, we set the out-of-plane moment in the field to be aligned along $c^*$ perpendicular to the $a$-$b$ plane, in accordance with the magnetisation and specific heat measurement setup where the field is applied along $c^*$. {\new By analysing the scattering patterns obtained from experiment and theoretical simulations both qualitatively and quantitatively, 
we believe that the exotic zero-up-zero-down-up-up order~($\circ$$\uparrow$$\circ$$\downarrow\uparrow\uparrow$)  as shown in Fig.~\ref{fig3}d is most likely to be the magnetic structure in the plateau phase~(see details in Extended Data Figs.~4 and 5 and Extended Data Table~1). The corresponding calculated magnetic structure factors are shown in the bottom of Fig.~\ref{fig3}c.} It is transitioning from the zigzag order ground state with four spins as a unit by partially flopping two of the six spins in an enlarged magnetic unit onto the honeycomb plane so that their components along the out-of-plane direction become zero.

Such a structure not only can produce the structure factors {\new that closely resemble} the experimental pattern as shown in the bottom of Fig.~\ref{fig3}c, but also naturally explains
the unexpected 1/3 magnetisation plateau in Fig.~\ref{fig2}a. First, the magnetic unit consisted of six magnetic spins can satisfy the necessary condition $n(S-m)=$ integer for
the 1/3 magnetisation plateau, where $n$ is the number of spins in a magnetic unit,
and $m$ is the fraction of the plateau\cite{PhysRevLett.78.1984}. Second, within one unit, the total spin $S=2$ is exactly 1/3 of $S=6$ in the fully polarised state. In fact, we have also attempted to examine the direction of the flopped spins in the plane. However, either parallel or antiparallel arrangement of the two flopped spins will lead to six peaks around the triangle centres in the second Brillouin zone exhibiting comparable intensities, which is in conflict with the dramatically different intensities of the two groups of peaks rotated by 180$^\circ$ in the experimental results. {\new We note that there are some slight discrepancies between the calculated and experimental patterns as shown in Fig.~\ref{fig3}c. We think there may be some disorder effect along the enlarged 6-site chain, and therefore which two of the six sites to have the zero moment along out of plane can have some uncertainties. Considering this may partially resolve the discrepancies. Furthermore, the fluctuations of the in-plane spins at the ``$\circ$" sites should also be considered~(see detailed discussions in Extended Data Fig.~4 and 5, and Extended Data Table~1).}



\begin{figure}[htb]
\centering
\includegraphics[width=3.4in]{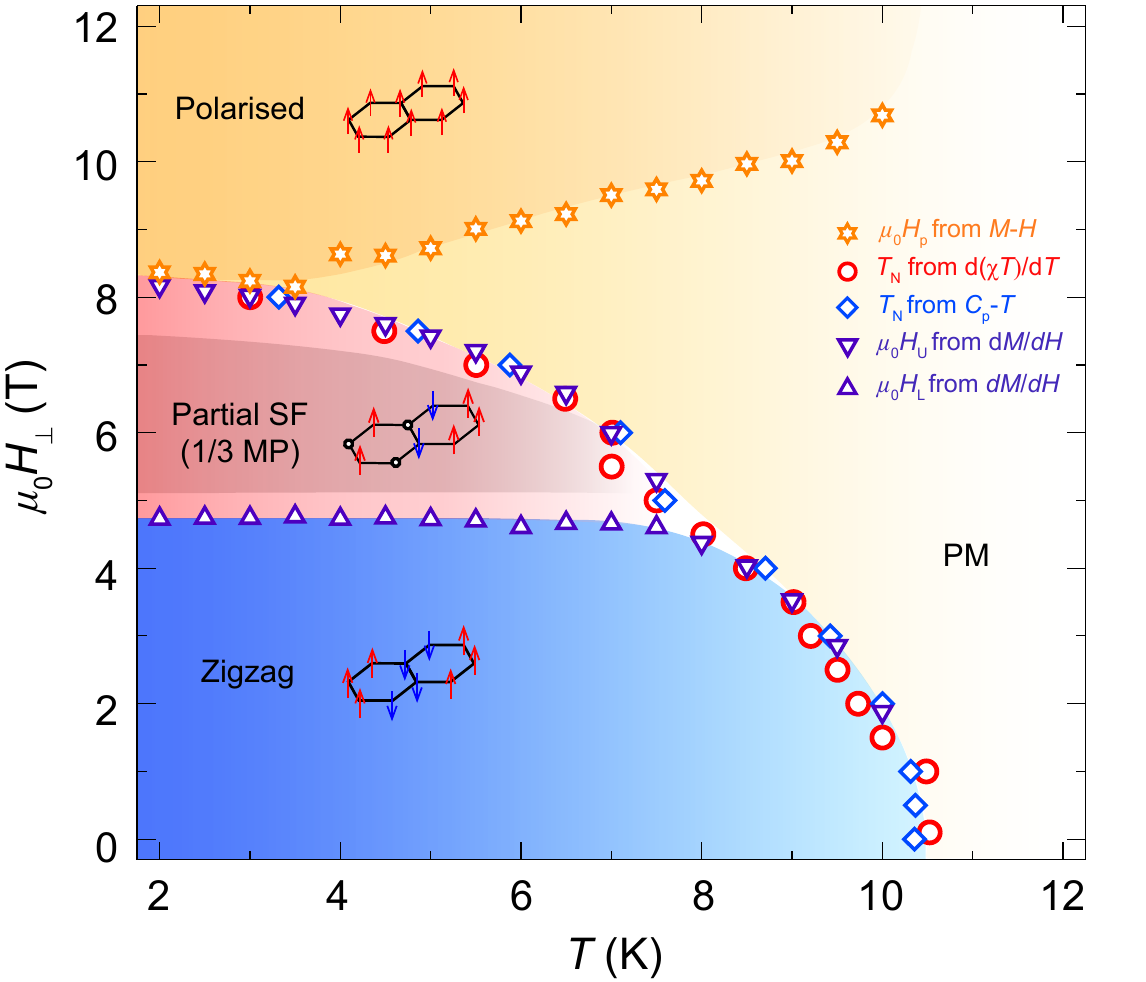}
\caption{\label{fig4}{{\bf Magnetic phase diagram of \NNBO with field applied perpendicular to the $a$-$b$ plane.} The lower blue zone denotes the zigzag antiferromagnetic state, and the central pink area represents the field-induced partial spin-flop (Partial SF) ferrimagnetic state, including a stable 1/3 magnetisation plateau phase (1/3 MP) and the intermediate transition states. The orange zone is the fully-polarised ferromagnetic state. PM denotes the paramagnetic state when the temperature is above $T_{\rm N}$. Orange stars represent the fully-polarised fields obtained from the $M$-$H$ curves. Red circles and blue diamonds denote the N\'{e}el temperatures extracted from the derivatives of the magnetic susceptibility (d$\chi T$/d$T$), and specific heat data, respectively. The violet up and down triangles represent the lower and upper critical fields of the 1/3 magnetisation plateau phase, obtained from d$M$/d$H$. The shade represents the stable 1/3 magnetisation plateau phase. Corresponding magnetic structures for the the various phases are sketched in the phase diagram. Symbols $\uparrow$, $\downarrow$, and $\circ$ represent respective spin component of 1, -1, and 0 along the out-of-plane direction.}}
\end{figure}

\medskip
\noindent{\bf Magnetic phase diagram}\\
By summarizing aforementioned thermodynamic and neutron scattering results, we obtain a magnetic phase diagram with $H\perp$ $a$-$b$ plane for \NNBO as shown in Fig.~\ref{fig4}. In the upper left corner, there is a fully-polarised state, whose phase boundary is determined from the $M$-$H$ curves in Fig.~\ref{fig2}a. For the high-temperature data where the saturation is not that obvious, we obtain the values of the fully-polarised field $\mu_{0}H_{\rm p}$ by taking the intersections of the linear fittings between the low- and high-field curves at different temperatures. The phase boundary between the long-range order and paramagnetic state is determined by extracting the peaks from the specific heat (Fig.~\ref{fig1}d) and derivatives of the susceptibility d$\chi T$/d$T$~(Extended Data Fig.~1d)\cite{PhysRevB.7.4197}, which are mutually consistent. The transition temperature is gradually suppressed with increasing magnetic field. More importantly, there is a 1/3 magnetisation plateau phase at low temperatures. Both the lower and upper boundaries of this phase can be determined by extracting the peaks from d$M$/d$H$ {\it vs.} $\mu_0H_{\perp}$ curves as shown in Fig.~\ref{fig2}b. The upper phase boundary set by d$M$/d$H$ overlaps with that set by d$\chi T$/d$T$ and the specific heat. The lower phase boundary remains almost flat until 8~K, where it merges into the outer boundary. It is worth mentioning that the upper-bound field at low temperatures and the transition field at 8~K and above coincide with the phase boundary between the long-range order and paramagnetic state. At low temperatures such as 2~K, the system undergoes two successive phase transitions from zigzag to plateau phase, and then to the fully-polarised state. As shown in Fig.~\ref{fig2}, the transitions under fields appear to be of second order with a finite transition width, making the plateau phase slightly narrower than the phase boundary determined by d$M$/d$H$.


\medskip
\noindent{\bf Crucial role of the Kitaev interaction}\\
Such an observation of the partial spin-flop transition induced 1/3 magnetisation plateau phase in a honeycomb-lattice antiferromagnet \NNBO is quite unusual. To stabilise such a phase, a key ingredient is the frustration\cite{Chubukov_1991,Honecker_1999,0034-4885-78-5-052502,PhysRevLett.85.3269,kawamura1985phase,PhysRevLett.62.2056,PhysRevLett.102.137201,PhysRevB.87.060407,PhysRevLett.112.127203}. For \NNBO with a honeycomb lattice, it does not have the geometrical frustration as  triangular and kagome lattices do. Although magnetic exchange frustration could be arising from the competing first- ($J_1$) and second-neighbour Heisenberg exchange couplings~($J_2$)~(Refs.~\cite{doi:10.1021/ic402131e,jacs131_8313,doi:10.1143/JPSJ.79.114705}),  and the $J_{1}$-$J_{2}$-$J_{3}$ (third-neighbour coupling) model can even give rise to the zigzag order\cite{epjb20_241}, it can hardly explain the highly anisotropic responses under in- and out-of-plane fields and thus the spin-flop transition observed in \NNBO. Intriguingly, taking into account the edge-shared octahedral structure similar to Kitaev materials such as $\alpha$-RuCl$_3$ and iridates\cite{0953-8984-29-49-493002,nrp1_264,PhysRevB.94.064435}, and strong spin-orbital coupling (SOC) in the vicinity of the heavy Bi atoms and strong Hund's coupling in Ni$^{2+}$, a strong Kitaev interaction, which is a natural source of the exchange frustration due to the bond-dependent anisotropy, may be realized in \NNBO~(Ref.~\cite{PhysRevLett.123.037203}). To examine the role of the Kitaev interaction in the exotic
magnetic behaviours observed above, we have performed
first-principle and many-body calculations (see Methods and Extended Data Fig.~6) for
the compound and {\new  provided evidence that the observed 1/3-plateau phase in Na$_3$Ni$_2$BiO$_6$ is selected by the Kitaev interaction via an intriguing mechanism that will be clarified below.}

\begin{figure}[htb]
\centering
\includegraphics[width=3.4in]{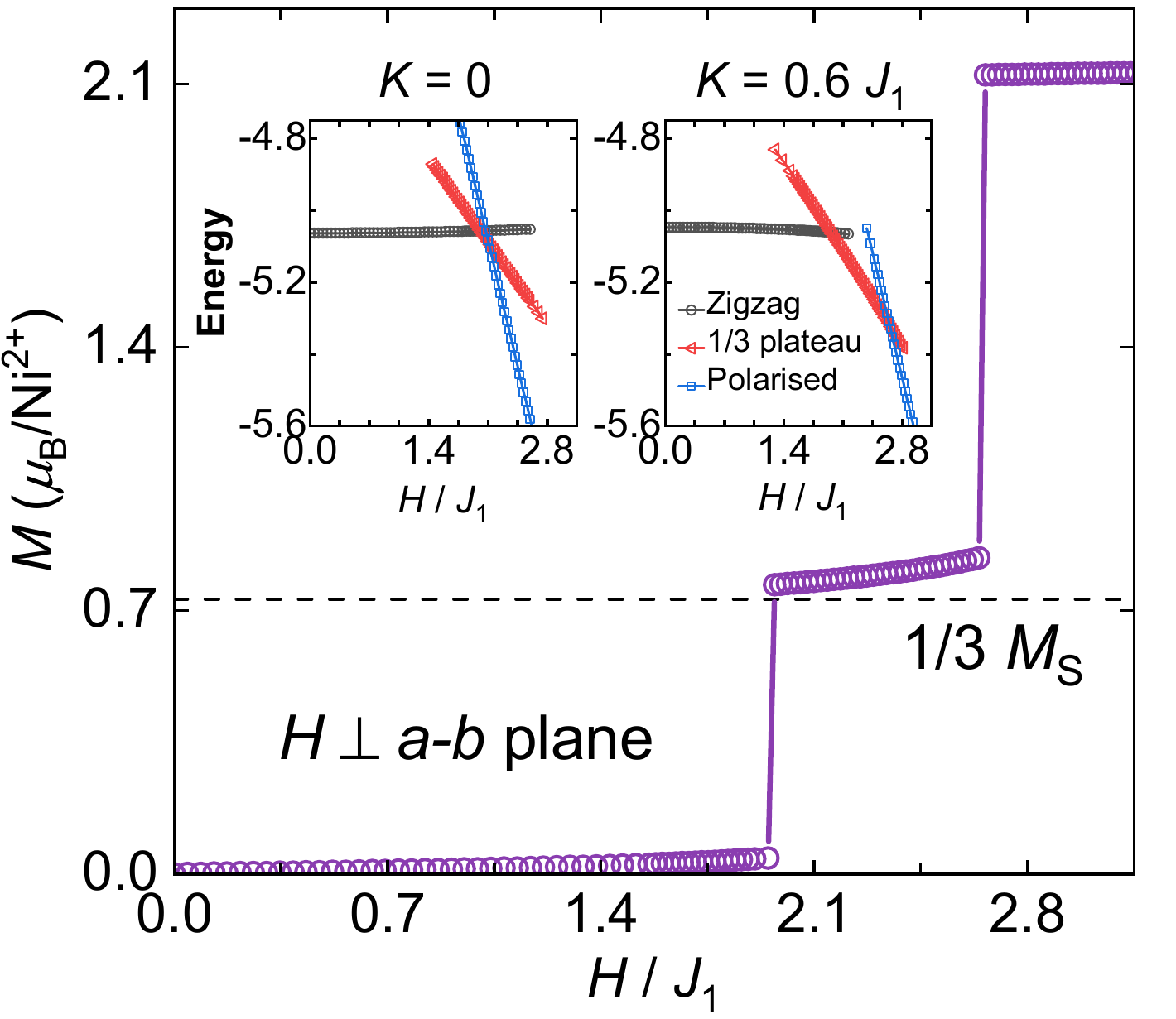}
\caption{\label{fig5}{\bf Magnetisation curve of the spin-1 $J_1$-$J_3$-$K$-$D$ model obtained by the tensor-network calculations.}
The magnetic field is applied along $c^*$,
i.e., perpendicular to the $a$-$b$ plane. There exists a 1/3
plateau in the ground-state magnetisation curve as shown in
the main figure, which is stabilised by the Kitaev interaction.
To explain this observation, we show the variational energy
of the zigzag, 1/3-plateau, and polarised states in two insets
with $K=0$ and $K=0.6$, respectively. Without the $K$ term,
energies of the zigzag, 1/3-plateau, and the polarised states
intersect virtually at the same field,  explaining the absence of
1/3-plateau structure in the $J_1$-$J_3$-$D$ model without $K$.
In contrast, when the Kitaev term is introduced to the model, the
1/3-plateau energy gets reduced and polarised energy increased.
The so-obtained 1/3-plateau, selected as the ground state by
anisotropic Kitaev interaction, has a finite slope in the
intermediate field regime resembling that of experimental
data in Fig.~\ref{fig2}a. The tensor-network calculations are
performed by retaining $\gamma=5$ bond states that is sufficient large
to compute accurately the magnetisation of the effective model for Na$_3$Ni$_2$BiO$_6$.}
\end{figure}

Our DFT calculations using the four-state method show that the third-nearest-neighbour exchange coupling {\new $J_3$ is comparable to the nearest-neighbour coupling $J_1$
but the second-nearest-neighbour coupling $J_2$}
is instead negligible. Considering the fact of easy-axis magnetic
anisotropy in the compound, it is naturally to add into the
spin Hamiltonian a single-ion anisotropy term $D$.
After a thorough scan of the possible models (see Extended
Data Table~3), we find that the Kitaev ($K$) term must be included
to reproduce the 1/3 plateau. Therefore, we take the $J_{1}$-$J_{3}$-$K$-$D$ model as a minimal model for the compound Na$_3$Ni$_2$BiO$_6$, which reads as
\begin{equation*}
\begin{split}
\hat{H}= J_{1}\sum_{\langle i,j\rangle} \mathbf{S}_{i}\cdot\mathbf{S}_{j}
+J_{3}\sum_{\langle\langle\langle i,j\rangle\rangle\rangle}\mathbf{S}_{i}
\cdot\mathbf{S}_{j} \\
+K\sum_{i,\alpha}\mathbf{S}_{i}^{\alpha}
\cdot\mathbf{S}_{i+\alpha}^{\alpha} +
D\sum_{\langle i\rangle}\mathbf({S}_{i}^{c})^2 {\new - H\sum_{\langle i\rangle} S_{i}^{c^*}},
\label{Eq:Haim}
\end{split}
\end{equation*}
where $\mathbf{S}_{i}$ is an $S = 1$ quantum spin operator at site $i$. Due to the slight lattice distortion, we set the single-ion anisotropic axis, {\it i.e.}, the quantized axis of $S_i^c$, as tilted about $10^{\circ}$ from the perpendicular $c^*$ axis. {\new The last term is the Zeeman term coupled to the $c^*$-axis ($i.e.$, out-of-plane) component of the quantum spin.} Taking $J_{1}$ as the energy scale, we find an effective $J_{1}$-$J_{3}$-$K$-$D$ model with the set of parameters ($J_{1}=-1$, $J_{3}=1$, $K=0.6$, $D=-3$) well explains the experimental observations.

In Fig.~\ref{fig5} we show the obtained magnetisation curve calculated from the spin-1 $J_{1}$-$J_{3}$-$K$-$D$ model, which indeed has a 1/3-plateau state under intermediate fields. To understand the role of Kitaev interaction, in the insets of Fig.~\ref{fig5} we show the variational energies of the zigzag, 1/3-plateau, and polarised states. The introduction of the $K$ term increases the exchange frustration and thus quantum fluctuations of the system, and lowers the variational energy of 1/3-plateau state while raising that of the polarised states, allowing the 1/3-plateau to appear in the intermediate field regime in Fig.~\ref{fig5}. {\new It is worth mentioning that while the coupling parameters in the minimal $J_{1}$-$J_{3}$-$K$-$D$ model have not been fine tuned to quantitatively fit the experimental results, we already find the ratio of the two critical fields $h_{c1}/h_{c2}\sim$1.4 matches that of the experiment. Therefore, the tensor-network calculations support that it is the Kitaev term that selects and stabilises the 1/3 plateau in \NNBO.} Furthermore, with this model we have also calculated the magnetisation with field parallel to the $a$-$b$ plane, and the results also match the experimental measurements~ (see Extended Data Fig.~1c). 

\medskip
\noindent{\new{\bf Discussions}}\\
{\new From the results above, we have demonstrated that there is a 1/3 magnetisation plateau phase in a honeycomb-lattice antiferromagnet \NNBO, proposed an exotic magnetic structure symbolised as $\circ$$\uparrow$$\circ$$\downarrow\uparrow\uparrow$ for the plateau phase, and revealed that the Kitaev interaction is indispensable to this phase. Nevertheless, the magnetic structure calculated from the $J_{1}$-$J_{3}$-$K$-$D$ model, however, is slightly different from the zero-up-zero-down-up-up ($\circ$$\uparrow$$\circ$$\downarrow\uparrow\uparrow$) configuration in Fig.~\ref{fig3}~d. It has the down-down-up-up-up-up ($\downarrow\downarrow\uparrow\uparrow\uparrow\uparrow$) as depicted in Extended Data Fig.~4e instead. We believe this discrepancy between theory and experiment is because our $J_{1}$-$J_{3}$-$K$-$D$ model may still be too condensed, without taking some other anisotropic terms such as the off-diagonal ones into account. In this sense, future inelastic neutron scattering studies on single crystals to parameterize these terms should be very interesting. Despite this small deficit, this effective minimal model captures all of the essence of the experimental measurements, including the zigzag order at zero field, spin-flop transitions~(Extended Data Table~3), and magnetisations under fields, in particular the 1/3 magnetisation plateau. With these, this work not only extends the study of the fractional magnetisation plateau phase to honeycomb-lattice compounds which conventionally do not exhibit geometrical frustrations, but also expands the territory of quantum magnets that host Kitaev physics from $S=1/2$ to higher-spin systems.}

\bigskip
\noindent {\bf Methods}\\
\noindent {\bf Single-crystal growth and characterisations.} High-quality single crystals of \NNBO were successfully grown by the flux method\cite{doi:10.1021/ic402131e}. The crystals were thin and transparent brown, with a typical mass of $\sim$7~mg for each piece. Magnetisation and specific heat measurements were conducted in a physical property measurement system PPMS-9T from Quantum Design. Later, we extended the magnetisation measurements up to 14~T in a PPMS-14T. Magnetisations were measured under magnetic fields both perpendicular and parallel to the $a$-$b$ plane with fixed-field sweeping-temperature and fixed-temperature sweeping-field modes. Specific heat measurements were carried out in the temperature region between 2 and 18~K in a series of magnetic fields applied perpendicular to the $a$-$b$ plane. Magnetisation was also measured as a function of field up to 14~T at High Magnetic Field Laboratory of the Chinese Academy of Sciences.


\bigskip
\noindent {\bf Neutron scattering experiments.} Neutron scattering measurements were performed on PELICAN, a cold-neutron time-of-flight spectrometer located at the OPAL of ANSTO in Australia\cite{yu2013pelican}. The sample array consisted of $\sim$140 pieces of single crystals weighing about 1~g in total. They were glued onto four rectangular aluminum plates by hydrogen-free Cytop grease and well coaligned using a backscattering Laue x-ray diffractometer. The aluminum plates were well tilted by 18.56$^{\circ}$ so that the horizontal plane was the ($H,\,K,\,0$) plane. The assembly with ($H,\,K,\,0$) as the horizontal plane was installed in a 7-T superconducting magnet. We set the angle where the [100] direction was parallel to the incident beam direction to be zero. Data were collected at 1.5 and 80~K with $E_{\rm i}= 3.70$~meV by rotating the sample about the vertical direction with a range of 360$^{\circ}$ in a 2$^{\circ}$ step. At 1.5~K, measurements were done in two magnetic fields of $\mu_0$$H$ = 0 and 7~T. For refinement purpose, additional neutron diffraction measurements on these single crystsals were carried out on AMATERAS, a cold-neutron time-of-flight spectrometer with a multiple-$E_{\rm i}$ option located at the MLF of J-PARC in Japan\cite{doi:10.1143/JPSJS.80SB.SB028}. We also performed powder neutron diffraction measurements on GPPD, a time-of-flight diffractometer located at CSNS in China. We adopted the monoclinic $C2/m$ structure with the refined lattice parameters in Ref.\cite{doi:10.1021/ic402131e}, with $a = 5.3998$~\AA, $b = 9.3518$~\AA, $c = 5.67997$~\AA, and $\beta = 108.56^\circ$. The wave vector $\bf Q$ is expressed as ($H,\,K,\,L$) in reciprocal lattice unit (r.l.u.) of $(a^*,\,b^*,\,c^*)$ = ($2\pi/a\cos\theta,\,2\pi/b,\, 2\pi/c\cos\theta$) with $\theta = 108.56^{\circ}-90^{\circ}=18.56^{\circ}$, representing the angle between $a(c)$ and $a^*(c^*)$. Since we tilted the $a$-$b$ plane by 18.56$^{\circ}$ from the horizontal plane, but applied a magnetic field of $\mu_{0}H = 7$~T along the vertical direction ($c$ axis), the component along the direction perpendicular to the $a$-$b$ plane ($c^*$) was $\sim$6.6~T, which pinned the system in the 1/3 magnetisation plateau phase.


\bigskip
\noindent {\bf Calculations of the magnetic structure factors.} The intensity of elastic neutron scattering was proportional to the square of the component of the magnetic structure factor,
\begin{equation*}\label{msf}
  \bm{I}(\mathbf{k})\propto\sum_{\alpha\beta}(\delta_{\alpha\beta}-\hat{k}_\alpha\hat{k}_\beta)\bf{M}^\alpha(\mathbf{k})M^\beta(\mathbf{k}),
\end{equation*}
where $\bf{M}(\bf{k})$ was the magnetic structure factor, the Fourier transform of the spin distribution with ${\bf M}({\bf k})=\sum_{\bf{r}_i} \bf{S}_{\bm{r}_i}{\rm e}^{i \bm{k}\cdot \bm{r}_i}$. ${\bf S}_{\bf{r}_i}$ of spin-1 had the value of 1, 0, and -1 for each site in the calculations. We tried the configurations of the honeycomb-lattice clusters with 6 and 24 sites in order to satisfy the period for the 1/3 magnetisation plateau phase. However, the corresponding magnetic structure factors would result in major Bragg peaks at $\mit\Gamma$ or $K$ points, which were not present in the experimental results. Reviewing the scattering patterns, the Bragg peaks at approximately $1/3$ and $2/3$ of the distance between $\mit\Gamma$ and $\mit\Gamma'$ implied a possible order with a 6-site chain possessing the translational symmetry instead of the 4-site magnetic unit of the zigzag order. Therefore, we calculated the magnetic structural factors based on a series of magnetic structures with a 6-site chain with translational symmetry.

\bigskip
\noindent {\bf First-principles and tensor-network calculations.}
The DFT calculations were performed by using the Vienna \emph{ab initio} simulation
package (VASP) with Perdew-Burke-Ernzerhof functional.
The experimental geometric structure\cite{doi:10.1021/ic402131e} was used and
the effective Coulomb interaction $U_{\rm eff} = 4$~eV and SOC were considered.
In the quantum many-body calculations of the spin-1 Kitaev-Heisenberg model, we
employed the infinite projected entangled pair state (iPEPS)\cite{Cirac2021RMP} and density matrix renormalization group (DMRG) approaches, and obtained accurate ground state properties.

\bigskip
\noindent {\bf Data availability}\\
\noindent Data supporting the findings of this study are available from the corresponding author J.S.W. (Email: jwen@nju.edu.cn) upon reasonable request.

\bigskip
\noindent {\bf Acknowledgements}\\
\noindent The work was supported by National Key Projects for Research and Development of China with Grant No.~2021YFA1400400, National Natural Science Foundation of China with Grant Nos.~12225407, 12074174, 12074175, 11904170, 11834014, 11974036, 12222412, 12004191 and 12204160, Natural Science Foundation of Jiangsu province with Grant Nos.~BK20190436 and BK20200738, China Postdoctoral Science Foundation with Grants No.~2022M711569 and 2022T150315, Jiangsu Province Excellent Postdoctoral Program with Grant No.~20220ZB5, Hubei Provincial Natural Science Foundation of China with Grant No.~2021CFB238, CAS Project for Young Scientists in Basic Research (YSBR-003), and Fundamental Research Funds for the Central Universities. We acknowledge the neutron beamtime from ANSTO with Proposal No.~P9334 and the great support from Gene Davidson in setting up and operation of the 7-T superconducting magnet, and the beamtime from J-PARC with Proposal No.~2022A0039. We would like to thank Yuyan~Han at High Magnetic Field Laboratory of the Chinese Academy of Sciences for assisting us in measuring the magnetisation under high magnetic fields.

\bigskip
\noindent {\bf Author contributions}\\
\noindent J.S.W. conceived the project. Y.Y.S.G. prepared the samples. Y.Y.S.G. carried out the magnetisation and specific heat measurements with assistance from S.Z. and F.Q.S. for the 14-T field measurements. Y.Y.S.G., S.B., D.H.Y., R.A.M, N.M., S.O.-K., L.H.H. and J.Z.H. performed the neutron scattering experiments. Y.Y.S.G., S.B. and J.S.W. analysed the experimental data. Z.-Y.D., N.X., Y.-P.G., Z.Q., Q.-B.Y., W.L., S.-L.Y. and J.-X.L. performed the theoretical calculations and analyses. J.S.W., Y.Y.S.G., W.L. and J.-X.L. wrote the paper with inputs from all co-authors.\\

\bigskip
\noindent {\bf Competing Interests}\\
\noindent The authors declare no competing financial interests.

\bigskip
\noindent {\bf Additional information}\\
\noindent Correspondence and request for materials should be addressed to J.S.W. ~(jwen@nju.edu.cn), J.-X.L.~(jxli@nju.edu.cn), S.-L.Y.~(slyu@nju.edu.cn) or W.L.~(w.li@itp.ac.cn).

\end{document}